\documentclass{evn2024}
\usepackage{graphicx}
\usepackage{amsfonts}
\usepackage{amsmath}
\usepackage{amssymb}
\begin{document}
   \title{MAD accretion and AGN jets -- an observational perspective}

   \author{T.~Savolainen\inst{1,2,3}
          \and
          W.~Chamani\inst{4}
          }

   \institute{
        Aalto University Department of Electronics and       Nanoengineering, PL 15500, FI--00076 Aalto, Finland
        \and
        Aalto University Mets\"ahovi Radio Observatory,       Mets\"ahovintie 114, FI--02540 Kylm\"al\"a, Finland
        \and
        Max Planck Institute for Radio Astronomy,
        Auf dem H\"ugel 69, D--53121 Bonn, Germany
        \and
        Deutsches GeoForschungsZentrum (GFZ), Telegrafenberg A17, 14473, Potsdam, Germany
             }

   \abstract{
    One of the major open questions related to the production of jets by accreting black holes is: why do sources with similar accretion powers produce so vastly different jet powers? What conditions are required to make a powerful jet? If jets are powered by the Blandford-Zjanek mechanism, two further parameters control the jet power besides the black hole mass -- black hole spin and the magnetic flux threading it. Since highly spinning black holes without jets appear to exist, the jet production efficiency may depend on whether the black hole managed to accrete high enough magnetic flux in the past. The highest-efficiency jets in this picture are launched from magnetically arrested disks (MADs). Here we discuss a method to test this hypothesis using VLBI core-shift measurements to estimate the jet magnetic flux.
   }

   \maketitle
%

\section{Introduction}

Understanding what controls the ejection of plasma jets from accreting black holes is a long-standing issue in relativistic astrophysics. The currently favoured theoretical picture uses magnetic fields as the means of tapping rotational energy either from the spinning black hole (Blandford \& Znajek\ \cite{blandford77}) or the inner accretion disk (Blandford \& Payne\ \cite{blandford82}) to power the jets. In recent years, general relativistic magnetohydrodynamic (GRMHD) simulations have provided support to this picture by demonstrating that a rotating black hole is indeed able to launch stable, high Lorentz factor jets via the Blandford-Znajek (BZ) mechanism if enough magnetic flux accumulates around the black hole either through advection from larger scales or in-situ generation by a dynamo effect (Tchekhovskoy et al.\ \cite{tchekhovskoy11}; Liska et al.\ \cite{liska20}).

One of the major open questions regarding black hole jet production is what creates the wide span of at least four orders of magnitude -- or possibly even a dichotomy -- in the jet power, $P_\mathrm{j}$, for sources with similar accretion disk luminosities and Eddington ratios (Sikora et al.\ \cite{sikora07}). Black hole mass and accretion rate, $\dot{M}$, alone cannot determine the jet production efficiency $p_\mathrm{j} = \frac{P_\mathrm{j}}{\dot{M} c^2}$, meaning that at least one more parameter is needed (Sikora et al.\ \cite{sikora07}; Gardner \& Done \cite{gardner14}, \cite{gardner17}). In the Blandford-Znajek-mechanism two additional parameters control the jet power: black hole spin and the magnetic flux threading the black hole. Hence, it is natural to explore the idea that variation of these two could cause the observed large range in the commonly used proxy for $p_\mathrm{j}$, the radio-loudness parameter $\mathcal{R}=L_5/L_B$, where $L_5$ and $L_B$ are the monochromatic luminosities at 5\,GHz and 4400\AA, respectively. Sources with $\mathcal{R} \lesssim 10$ are considered radio-quiet.

One possibility is that the wide distribution of $\mathcal{R}$ is solely due to the spin distribution of supermassive black holes (SMBH). Tchekhovskoy et al. (\cite{tchekhovskoy10}) showed that it is possible to get jet power variations of $\sim 1000$ from realistic SMBH spin distributions. On the other hand, the existence of radio-quiet AGN with measured high spins does not fit well in this model (Reynolds \cite{reynolds21}). However, the current spin measurements are model-dependent and can have significant unaccounted systematics. If the real geometry of the observed AGN cores differs from the assumptions of the relativistic reflection model, high spin could still be a necessary and sufficient condition for radio-loudness (e.g., Miller \& Turner \cite{miller13}, Done \& Jin \cite{done16}). 

Another possible parameter controlling the jet production efficiency is the magnetic flux threading the black hole. Sikora \& Begelman (\cite{sikora13}) have proposed that only those (high spin) sources that manage to accrete enough magnetic flux to develop a magnetically arrested disk (MAD; Narayan et al. \cite{narayan03}, Tchekhovskoy et al. \cite{tchekhovskoy11}), will produce high-power jets and magnetic flux threading the black hole would thus be the dominant parameter controlling the radio-loudness of an AGN. Consequently, radio-loudness would depend on the source's accretion history, since the poloidal magnetic field transport likely depends on the details of the accretion process (see e.g., Begelman \cite{begelman24}). Sikora \& Begelman (\cite{sikora13}) call this the "magnetic flux paradigm" as opposed to the "spin paradigm". While both parameters, magnetic flux and black hole spin, likely play a role in setting the jet power, these "paradigms" refer to the \emph{dominant} factor in determining the radio-loudness of an AGN. 

In this paper, we discuss how to constrain the black hole magnetic flux over a wide range of $\mathcal{R}$ to investigate magnetic fields' role in setting jet production efficiency. In Sect.~2, we briefly review the observational evidence for the high-$p_\mathrm{j}$ jets being launched from MAD systems and in Sect.~3, we present a method to estimate the BH magnetic flux in sources with low-$p_\mathrm{j}$ jets. 


\section{Evidence for powerful jets being related to MAD accretion}

The broadband spectral properties of self-absorbed synchrotron emission depend on the magnetic field strength in the plasma. This makes it possible to use multi-frequency VLBI observations to constrain the magnetic field strength in a jet by either measuring the spectral turnover of a resolved emission region or measuring the frequency-dependent shift of the location of the synchrotron photosphere along the jet (e.g., Savolainen et al. \cite{savolainen08}, Lobanov \cite{lobanov98}, Chamani et al. \cite{chamani23}). Pushkarev et al. (\cite{pushkarev12}) applied the latter, the so-called core-shift method to measure equipartition magnetic field strengths in 89 blazar jets and Zamaninasab et al. (\cite{zamaninasab14}) used these measurements together with magnetic field strengths from several studies of individual radio galaxies to estimate the poloidal magnetic flux in jets over a wide range of accretion disk luminosities, and hence, accretion rates. By assuming flux freezing conditions, the jet poloidal magnetic flux, $\Phi_\mathrm{jet}$ can be equated with the magnetic flux threading the black hole, $\Phi_\mathrm{BH}$. Zamaninasab et al. (\cite{zamaninasab14}) compared $\Phi_\mathrm{jet}$ with the prediction from the GRMHD simulations that the saturation value for the magnetic flux threading the black hole in a MAD state is 
\begin{equation}
\Phi_\mathrm{BH} \sim 50 (\dot{M} r_g^2 c)^{1/2},    
\end{equation}
where $r_g$ is the gravitational radius of the black hole and $c$ is the speed of light (Tchekhovskoy et al. \cite{tchekhovskoy11}). The results by Zamaninasab et al. (\cite{zamaninasab14}) show that the numerical factor in Eq.~1 is indeed $52\pm5$, thus fully agreeing with these sources being in a MAD state. This result has been confirmed by other groups who have relaxed some of the assumptions in the original analysis (e.g., Mocz \& Guo \cite{mocz15}, Zdziarski et al. \cite{zdziarski15}). There is thus observational evidence indicating that the AGN with high jet production efficiency\footnote{By high-efficiency jets, we mean jets with powers similar to or higher than the accretion power.} are likely to be in a MAD state. 

In addition to core-shift-based analyses, there exists also other evidence for strong magnetic fields in jetted AGN. This includes event horizon scale polarization VLBI measurements of the black hole M\,87*, which, when combined with a conservative lower limit on the jet power in M\,87, favour strongly magnetized MAD models (EHT Collaboration \cite{ehtc21}). Assuming BZ-mechanism, Kino et al. (\cite{kino22}) used the measured velocity profile of the M\,87 jet to constrain the horizon scale magnetic field strength and again obtained values consistent with MAD accretion. Recently, a far-infrared polarimetry study of dust cores in radio-loud and radio-quiet AGN showed different polarization properties between the two, with radio-loud sources having higher degrees of linear polarization of dust emission (Lopez-Rodriguez et al. \cite{lopez23}). Since the far-infrared polarization arises mainly from magnetically aligned dust grains at a scale of 5$-$130\,pc, these results indicate that the jet power and the strength of the magnetic field surrounding the AGN are correlated. 

Thus, multiple lines of evidence now suggest that AGN with high-efficiency jets have strong -- often saturation level -- magnetic fields around the black hole. However, this does not necessarily mean that the magnetic field \emph{is} the parameter controlling the jet production efficiency. To answer the question about the driving force behind the broad radio-loudness distribution for sources with similar accretion powers, we need a way to measure the magnetic flux in radio-quiet sources. If the sources with low $p_\mathrm{j}$ consistently have $\Phi_\mathrm{BH}$ values that are well below the MAD levels, this would point towards the magnetic flux paradigm of the AGN jet production efficiency (Sikora \& Begelman \cite{sikora13}). 


\section{New test for the magnetic flux paradigm}

\begin{figure}
   \centering
   \includegraphics[width=4.25cm]{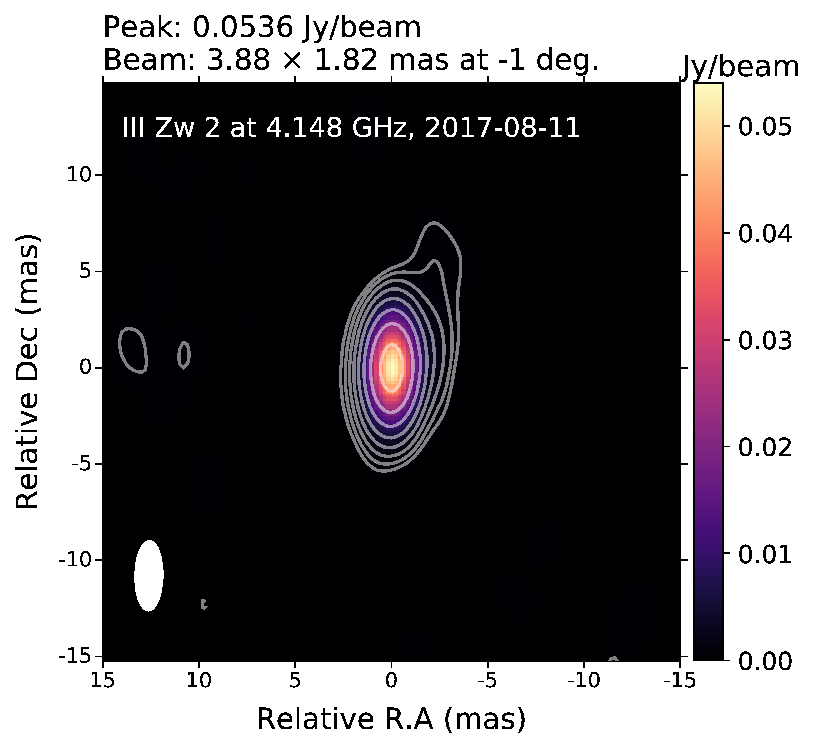}
   \includegraphics[width=4.2cm]{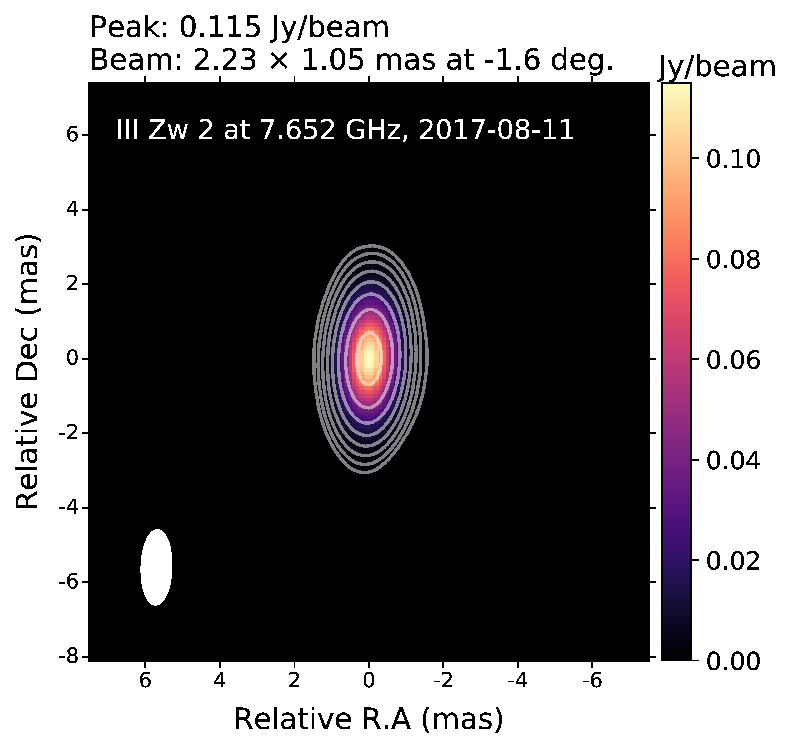} \\
   \includegraphics[width=4.2cm]{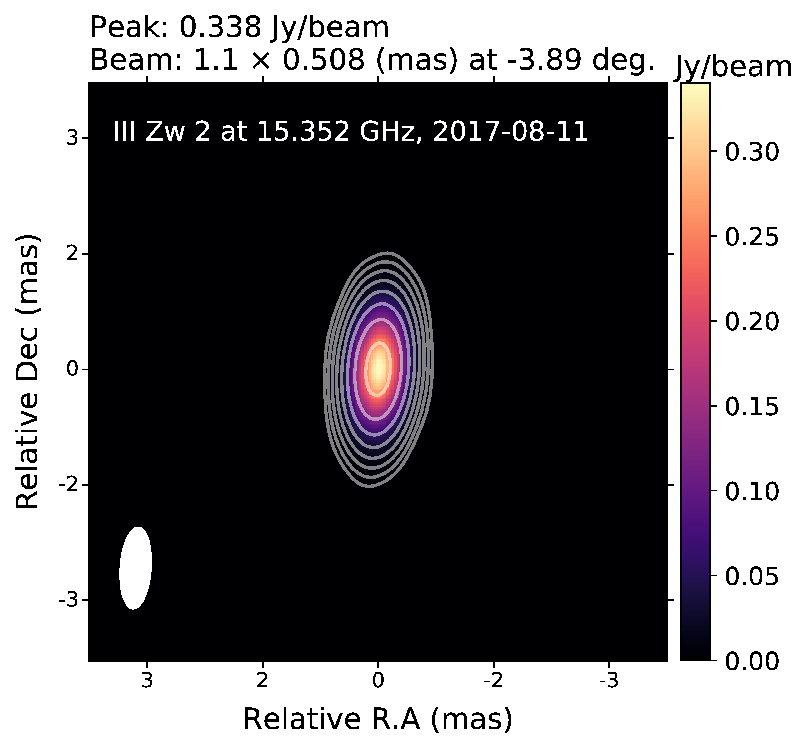}
   \includegraphics[width=4.25cm]{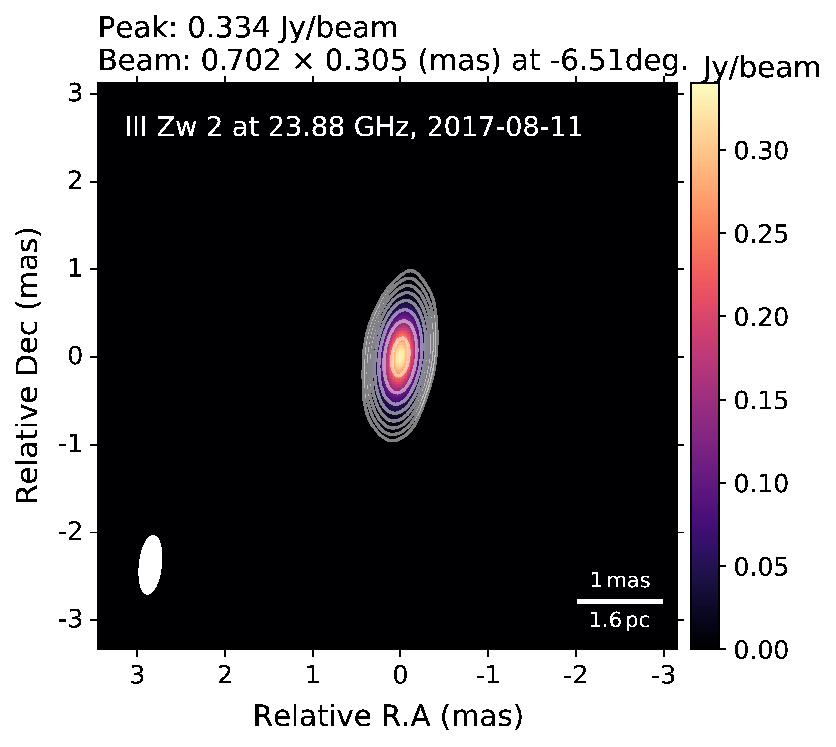}
   \caption{Radio-intermediate quasar III\,Zw\,2 observed quasi-simultaneously at 4.1\,GHz (top left), 7.7\,GHz (top right), 15.4\,GHz (bottom left) and 23.9\,GHz (bottom right) with the VLBA. In addition to III\,Zw\,2, three phase-reference calibrators were also observed to allow core shift measurements using relative astrometry. Credit: Chamani et al. (\cite{chamani21}), reproduced with permission \copyright ESO.  \label{fig:IIIZw2} }
\end{figure}

\begin{figure*}
    \centering
    \includegraphics[width=1.0\columnwidth]{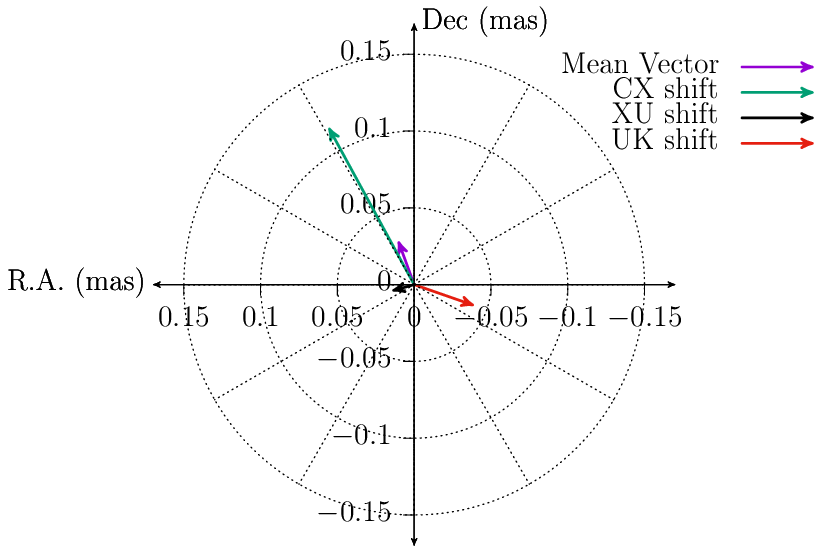}
    \includegraphics[width=1.0\columnwidth]{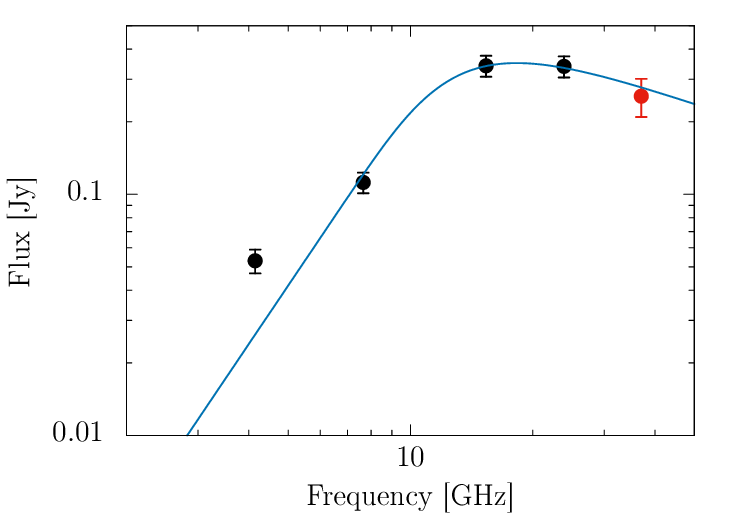}
    \caption{(Left:) Results of the core shift measurement for III\,Zw\,2. The plot shows core shift vectors for 4$-$8\,GHz (CX), 8$-$15\,GHz (XU), and 15$-$24\,GHz (UK). The 4$-$24\,GHz core shift is not significant, and we derive an upper limit of 0.16\,mas ($1\sigma$). (Right:) Core spectrum of III\,Zw\,2 between 4 and 24\,GHz (black points) and single-dish flux density at 37\,GHz (red point). Black line shows the fitted self-absorbed synchrotron spectrum. Credit: Chamani et al. (\cite{chamani21}), reproduced with permission \copyright ESO. \label{fig:coreshift} }
\end{figure*}

Since magnetic flux threading the black hole is difficult to directly probe beyond M\,87* (EHT Collaboration \cite{ehtc21}) and Sgr~A* (EHT Collaboration \cite{ehtc24}), our best bet probably is finding a sample of low-$\mathcal{R}$ (i.e., low-$p_\mathrm{j}$) AGN that still have detectable jets that allow constraining the magnetic field strengths via multi-frequency VLBI observations and then use their inferred $\Phi_\mathrm{jet}$ as a proxy for $\Phi_\mathrm{BH}$. Luckily, such radio-quiet sources with jets exist (e.g., Blundell et al. \cite{blundell96}, Ulvestad et al. \cite{ulvestad05}). Furthermore, so-called \emph{radio-intermediate} quasars are even better suited for the task. These sources with $10 \lesssim \mathcal{R} \lesssim 100$ are thought to be relativistically boosted counterparts of radio-quiet quasars (Falcke et al. \cite{falcke96a}, \cite{falcke96b}, Blundell \& Beasley \cite{blundell98}). Since their milliarcsecond radio emission is brighter than in radio-quiet quasars, the resulting higher signal-to-noise ratio in VLBI observations provides both higher astrometric accuracy in core-shift measurements and tighter constraints on the source size, and consequently better estimates on the magnetic field strength.

We have observed a pilot sample of four radio-quiet and radio-intermediate quasars with the Very Long Baseline Array (VLBA) at 4, 8, 15, and 24\,GHz using phase-referencing for relative astrometry. The aim is to measure or put tight upper limits to the core shift in these low-$p_\mathrm{j}$ jets to constrain their magnetic flux (Lobanov \cite{lobanov98}, Zamaninasab et al. \cite{zamaninasab14}). Furthermore, measuring the synchrotron turnover of a partially resolved core can give further constraints on the jet magnetic flux without needing to assume equipartition. The sources in the sample are radio-intermediate quasars III\,Zw\,2 (Mrk\,1501, PG\,0007+106), PG\,1309+355, and PG\,2209+184 and a radio-quiet quasar H\,1821+643. 

   \begin{figure}
   \centering
   \includegraphics[trim={1cm 0 1cm 0},clip,width=9.5cm]{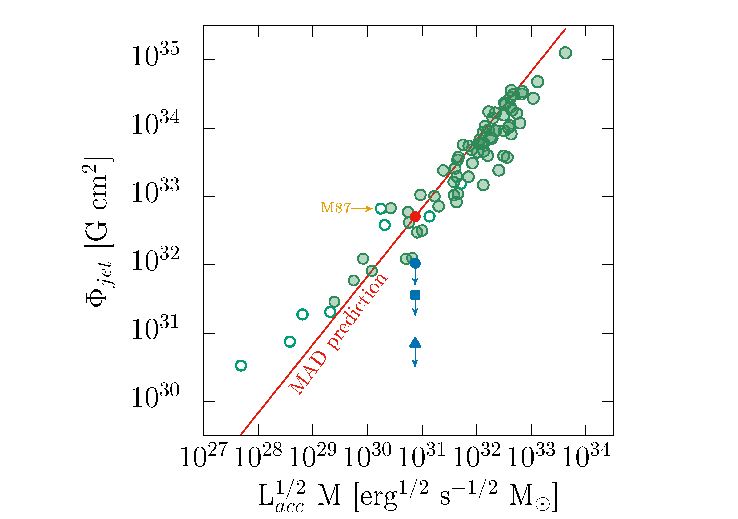}
      \caption{Jet magnetic flux as a function of $L_\mathrm{acc} M$, where $L_\mathrm{acc}$ is the accretion disk luminosity and $M$ is the mass of the black hole. Filled and open green data points are blazars and radio galaxies, respectively, from Zamaninasab et al. (\cite{zamaninasab14}). Blue points are measured upper limits for III\,Zw\,2 from the core-shift assuming equipartition (circle), from the core-shift without assuming equipartition (triangle), and from the spectral turnover of the synchrotron self-absorbed core (square). The red line shows the prediction for the MAD state. Credit: Chamani et al. (\cite{chamani21}), reproduced with permission \copyright ESO. \label{fig:MAD} }
   \end{figure}
%

In Chamani et al. (\cite{chamani21}), we presented the analysis of the observations of III\,Zw\,2. The multi-frequency images in Fig.~\ref{fig:IIIZw2} show this source to be very compact at all the frequencies and due to the compactness, it was impossible to use the self-referencing technique (e.g., Pushkarev et al. \cite{pushkarev12}) to measure the core shift. Instead, three nearby phase calibrator sources were observed to allow core shift measurement via relative astrometry. Since the phase calibrators did have extended structures, we used the self-referencing technique for them and then combined the resulting calibrator core shifts with the relative astrometry between them and III\,Zw\,2. The final core shift for III\,Zw\,2 was not significant given the astrometric uncertainties, but we could place an upper limit of 0.16\,mas for the 4$-$24\,GHz core shift in III\,Zw\,2 (left panel of Fig.~\ref{fig:coreshift}). This upper limit was then used to derive an upper limit for the magnetic field strength at 1\,pc from the jet apex, $B_\mathrm{1pc} \lesssim 60$\,mG assuming equipartition and $B_\mathrm{1pc} \lesssim 4$\,mG without assuming equipartition (Chamani et al. \cite{chamani21}). The estimated upper limit for the jet magnetic flux is a factor of five below the MAD level $\Phi_\mathrm{BH}$ in the equipartition case and a factor of about 70 below the MAD level $\Phi_\mathrm{BH}$ in the non-equipartition case (Fig.~\ref{fig:MAD}).

As shown in the right panel of Fig.~\ref{fig:coreshift}, the core spectrum of III\,Zw\,2 is self-absorbed and shows the spectral turnover in our measurements. This gives a possibility to infer the magnetic field strength following Marscher (\cite{marscher83}) and Savolainen et al. (\cite{savolainen08}) without the need to assume equipartition. Since the source is only marginally resolved, we could again only derive an upper limit from this analysis, $B_\mathrm{SSA} \lesssim 20$\,GHz. This leads to a jet magnetic flux upper limit that is a factor of about 15 below the MAD level (Fig.~\ref{fig:MAD}; Chamani et al. \cite{chamani21}).

The above results show that the radio-intermediate quasar III\,Zw\,2 has not reached the MAD level magnetic flux. On the other hand, fitting our combined XMM-\emph{Newton} and NuSTAR X-ray spectroscopy data with a disk reflection model results in almost maximal value for the black hole spin in III\,Zw\,2 (Chamani et al. \cite{chamani20}). While the signal-to-noise ratio of the X-ray data is not very high and we will likely have to wait for the next-generation high-throughput X-ray observatories like \emph{Athena} to have a more robust spin measurement, the combination of the low magnetic flux and the possibly high BH spin in this radio-intermediate source is fully compatible with the magnetic flux paradigm of Sikora \& Begelman (\cite{sikora13}). 

\section{Conclusions}

Observational evidence indicates that AGN with high-efficiency jets are often MADs. This is compatible with the idea that the magnetic flux threading the black hole is the parameter controlling jet production efficiency and consequently the radio-loudness of the source. To properly test this idea, we need to infer the black hole magnetic flux for a sample of sources that have a range of jet production efficiencies. This is challenging since our current best way to infer $\Phi_\mathrm{BH}$ is to use the jet magnetic flux inferred from multi-frequency VLBI observations as a proxy. The consequence is that the measurement becomes increasingly difficult as the radio-loudness of the sources decreases. We have, however, demonstrated that it is possible to find radio-quiet and especially radio-intermediate AGN for which this analysis can be done and the first results from our VLBA pilot program show that at least in III\,Zw\,2 the jet magnetic flux is clearly sub-MAD. 

With the current capabilities of the VLBA, it will be possible to moderately expand our sample, but the ngVLA will be a real game-changer in this regard allowing accurate multi-frequency astrometry for much weaker source samples. Combined with significantly improved X-ray spectroscopy from \emph{Athena} for black hole spin estimation, this holds a promise of solving the decades-old question: why is there a four orders of magnitude range in jet powers for sources with similar accretion powers?

\begin{acknowledgements}
This work was partly supported by the Research Council of Finland grant 362572.
The Very Long Baseline Array and the Green Bank Telescope is/are operated by The National Radio Astronomy Observatory, a facility of the National Science Foundation operated under a cooperative agreement by Associated Universities, Inc.
This work made use of the Swinburne University of Technology software correlator, developed as part of the Australian Major National Research Facilities Programme and operated under licence.

\end{acknowledgements}

\end{document}